\documentclass[preprint,aps,12pt,showpacs,nofootinbib,tightenlines]{revtex4}
\usepackage{amsmath}
\usepackage{amssymb}
\usepackage{epsfig}
\usepackage{graphicx}
\begin{document}
%%%%%%%%%%%%%%%%%%%%%%%%%%%%%%%%%%%%%%%%%%%%%
%%%%%%%%%%%%%%%%%%%%%%%%%%%%%%%%%%%%%%%%%%%%%%%%%%%%%%
\def\pslash{\rlap{\hspace{0.02cm}/}{p}}
\def\eslash{\rlap{\hspace{0.02cm}/}{e}}
%%%%%%%%%%%%%%%%%%%%%%%%%%%%%%%%%%%%%%%%%%%%%%%%%%%%%%
\title{Probing the lightest new gauge boson $B_H$ in the littlest Higgs model
via the processes $\gamma\gamma \rightarrow f\bar{f}B_H$ at the
ILC}
\author{Xuelei Wang} \email{wangxuelei@sina.com}
\author{Suzhen Liu}
\author{Qingguo Zeng}
\author{Zhenlan Jin}
\affiliation{College of Physics and Information Engineering, Henan
Normal University, Xinxiang, Henan 453007. P.R. China}
\thanks{This work is supported by the National Natural Science
Foundation of China(Grant No.10375017 and No.10575029).}

\date{\today}
\begin{abstract}

\indent  The neutral gauge boson $B_H$ with the mass of hundreds
GeV, is the lightest particle predicted by
 the littlest Higgs(LH) model, and such particle should be the
 first signal of the LH model at the planed ILC if it exists indeed.
In this paper, we study some processes of the $B_H$ production
associated with the fermion pair at the ILC, i.e.,
$\gamma\gamma\rightarrow f\bar{f}B_{H}$. The studies show that the
most promising processes to detect $B_H$ among
$\gamma\gamma\rightarrow f\bar{f}B_{H}$ are
$\gamma\gamma\rightarrow l'^+l'^-B_{H}(l'=e,\mu)$, and they can
produce the sufficient signals in most parameter space preferred
by the electroweak precision data at the ILC. On the other hand,
the signal produced via the certain $B_H$ decay modes is typical
and such signal can be easily identified from the SM background.
Therefore, $B_H$, the lightest gauge boson in the LH model would
be detectable at the photon collider realized at the ILC.
\end{abstract}


\maketitle
\newpage

\section{Introduction}
The little Higgs
models\cite{LH,little-1,little-2,little-3,littlest}
 were recently proposed to solve the hierarchy problem in the standard model(SM)
 by protecting the Higgs mass from
quadratical divergences at one-loop order, and thus can be
regarded as one of the important candidates of the new physics
beyond the SM. The key feature of this kind of models is that the
Higgs boson is a pseudo-Goldstone boson of a global symmetry
breaking at a scale $\Lambda\sim$ 10TeV, so that the Higgs boson
mass is naturally light. The light Higgs boson mass is protected
from the one-loop quadratic divergences by introducing a few new
particles with the same statistics as the corresponding SM
particles.

Among various little Higgs models, the littlest Higgs(LH)
model\cite{littlest} is a simplest and phenomenologically viable
one to realize the little Higgs idea. In this model, new charged
heavy vector bosons $W^{\pm}_{H}$, neutral heavy vector bosons
$Z_{H},B_{H}$, a heavy vector-like quark T and charged or neutral
heavy Higgs scalars are present which just cancel the quadratic
divergences by the SM gauge boson loops, the top quark loop and
Higgs self-interaction, respectively. These new particles might
produce characteristic signatures at present and future high
energy collider experiments\cite{ signatures-LH, Han,LHC}. In the
literatures\cite{Han,LHC}, the phenomenologies of the LH model at
the LHC have been  studied, showing that the LHC has the potential
to detect these particles. In the LH model, however, we find that
the globe symmetry structure $SU(5)/SO(5)$ allows a substantially
light $B_H$, light enough to be produced on shell at a 500 GeV
linear collider. $B_H$, as the lightest new particle in the LH
model, would play an important role in the phenomenological
studies of the LH model. Such gauge boson can be probed indirectly
through its contributions to some
processes\cite{Conley,correction-BH}. On the other hand, such
particle would be the first signal of the LH model at high energy
experiments, and the direct detection of it can provide a robust
evidence of the model. Although the LHC has the considerable
potential to detect $B_H$\cite{Han,LHC}, the detailed study of its
properties needs the precision measurement at the future high
energy and luminosity linear collider, and such work will be
performed at the planned International Linear Collider(ILC), with
the center of mass(c.m) energy $\sqrt{s}$ =300 GeV-1.5 TeV and the
integrated luminosity 500 $fb^{-1}$ within the first four year
running\cite{ILC}. The gauge boson $B_H$ would be light enough to
be produced at the first running of the ILC. So people also pay
much attention to study the $B_H$ production mechanism at the ILC.
Some $B_H$ production processes via $e^{+}e^{-}$ collision at the
ILC have been done \cite{e-production}. A unique feature of the
ILC is that it can be transformed to $\gamma\gamma$ or $e\gamma$
collisions with the photon beams generated by using the Compton
backscattering of the initial electron and laser beams. In this
case, the energy and luminosity of the photon beams would be the
same order of magnitude of the original electron beams, and the
set of final states at a photon collider is much richer than that
in the $e^+e^-$ mode. So the realization of the photon collider
will open a wider window to probe $B_{H}$. Some $B_H$ production
processes at the photon collider have also been
studied\cite{r-production}. In this paper, we study other
interesting $B_{H}$ production processes at the photon collider,
i.e., $\gamma\gamma\rightarrow f\bar{f}B_H.$ Here $f$ represent
all fermions in the SM. Our results show that the productions of
$B_H$ associated with $e^+e^-$ or $\mu^+\mu^-$ can also provide an
ideal way to probe $B_H$ with clean background.

This paper is organized as follows. In Sec. II, we first present
the key idea of the LH model and summarize some couplings in the
LH model related to our study, then we give our calculations of
the cross sections for the processes $\gamma\gamma\rightarrow
f\bar{f}B_{H}$. The numerical results and conclusions will be
shown in Sec. III.

\section{The key idea of the LH model and the cross sections of the processes $\gamma\gamma\rightarrow f\overline{f}B_H$}

The LH model is the most economical one among various little Higgs
models. It is based on a non-linear sigma model and consists of a
global $SU(5)$ symmetry which is broken down to $SO(5)$ by a
vacuum condensate $f\sim\frac{\Lambda_{s}}{4\pi}\sim$ TeV. Such
breaking scenario results in 14 Goldstone bosons. Four of them are
eaten by the broken gauge generators and will become the
longitudinal modes of four new massive gauge bosons, leaving 10
states that transform under the SM gauge group as a complex Higgs
doublet and a complex scalar triplet. A subgroup $[SU(2)\times
U(1)]^{2}$ of the global $SU(5)$ which is gauged with gauge
couplings $g_{1},g_{2},g'_{1},g'_{2}$, respectively, is
spontaneously broken down to its diagonal $SU(2)_L\times U(1)_Y$
subgroup. Such diagonal group is identified as the SM electroweak
gauge group and the mass eigenstates of the gauge bosons after the
symmetry breaking are
\begin{eqnarray}
W=sW_1+cW_2,~~~~W'=-cW_1+sW_2,\\ \nonumber
B=s'B_1+c'B_2,~~~~B'=-c'B_1+s'B_2.
\end{eqnarray}
Where the $W,B$ are the massless gauge bosons associated with the
generators of the electroweak gauge group $SU(2)_L\times U(1)_Y$.
The $W'$ and $B'$ are the massive gauge bosons associated with the
four broken generators of $[SU(2)\times U(1)]^2$. Using the mixing
parameters $c(s=\sqrt{1-c^2})$ and $c'(s'=\sqrt{1-c'^2})$, one can
represent the SM gauge coupling constants as $g=g_1s=g_2c$ and
$g'= g'_1s' = g'_2c'$.

  After electroweak symmetry breaking, the final observed mass eigenstates are obtained via mixing
between the heavy($W'$, $B'$) and light ($W$, $B$) gauge bosons.
They include the light SM-like bosons $W_L^{\pm},Z_L$ and $A_L$
observed at experiments, and new heavy bosons $W^{\pm}_H,Z_H$ and
$B_H$ that could be observed in future experiments. The masses of
neutral gauge bosons are given to $O(v^2/f^2)$ by \cite{Han,
Conley}
\begin{eqnarray}
M^{2}_{A_{L}}&=&0, \\ \nonumber
M^{2}_{B_{H}}&=&(M^{SM}_{Z})^{2}s^{2}_{W}\{\frac{f^{2}}{5s'^2c'^2v^2}-1
+\frac{v^2}{2f^2}[\frac{5(c'^2-s'^2)^2}{2s^2_W}-\chi_H\frac{g}{g'}\frac{c'^2s^2+c^2s'^2}{cc'ss'}]\},
\\ \nonumber
M^{2}_{Z_{L}}&=&(M^{SM}_{Z})^{2}\{1-\frac{v^{2}}{f^{2}}[\frac{1}{6}+\frac{1}{4}(c^{2}-s^{2})^{2}+
\frac{5}{4}(c'^{2}-s'^{2})^{2}]+8\frac{v'^2}{v^2}\},
\\ \nonumber
M^{2}_{Z_{H}}&=(&M^{SM}_{W})^{2}\{\frac{f^2}{s^2c^2v^2}-1+\frac{v^2}{2f^2}[\frac{(c^2-s^2)^2}{2c_W^2}
+\chi_H\frac{g'}{g}\frac{c'^2s^2+c^2s'^2}{cc'ss'}]\}.
\end{eqnarray}
Where,
$\chi_{H}=\frac{5}{2}gg'\frac{scs'c'(c^{2}s'^{2}+s^{2}c'^{2})}{5g^{2}s'^{2}c'^{2}-g'^2s^{2}c^{2}}$,
 $s_{W}(c_{W})$ represents the sine(cosine) of the weak mixing
angle, $v$=246 GeV is the elecroweak scale and $v'$ is the vev of
the scalar $SU(2)_{L}$ triplet.

The effective non-linear lagrangian invariant under the local
gauge group $[SU(2)\times U(1)]^{2}$ can be written as
\begin{eqnarray}
 \mathcal{L}_{eff}=\mathcal{L}_{G}+\mathcal{L}_{F}+\mathcal{L}_{\Sigma}
 +\mathcal{L}_{Y}-V_{CW}(\Sigma).
\end{eqnarray}
 $\mathcal{L}_{G}$ consists of the pure gauge terms;
  $\mathcal{L}_{F}$ is the fermion kinetic terms,
  $\mathcal{L}_{\Sigma}$ consists of the $\sigma$-model terms of the LH
model, $\mathcal{L}_{Y}$ is the Yukawa couplings of fermions and
pseudo-Goldstone bosons, and $V_{CW}(\Sigma)$ is the
Coleman-Weinberg potential generated radiatively
 from $\mathcal{L}_{Y}$ and $\mathcal{L}_{\Sigma}$. The couplings
 related to our work are the couplings of the neutral gauge bosons with
 fermion pair which are included in $\mathcal{L}_{F}$. Such fermion kinetic
 terms take the generic form
\begin{eqnarray}
 \mathcal{L}_{F}=\sum_{f}\bar{\psi}_fi\gamma^{\mu}D_{\mu}\psi_f ,
\end{eqnarray}
with
\begin{eqnarray}
 D_{\mu}=\partial_{\mu}-i\sum^2_{j=1}(g_jW_{j\mu}+g'_jB_{j\mu}).\nonumber
\end{eqnarray}

The couplings of the neutral gauge bosons $V_i$ with fermion pair
can be written in form of
$V_{\mu}^{V_if\bar{f}}=i\gamma_{\mu}g^{V_if\bar{f}}=i\gamma_{\mu}(g_V^{V_if\bar{f}}+g_A^{V_if\bar{f}}\gamma^5)$,
and the explicit expressions of the couplings related with $B_H$
are \cite{Han}
\begin{eqnarray}
g^{B_{H}l^{+}_il^{-}_i}_{V}&=&\frac{g'}{2s'c'}(2y_e-\frac{9}{5}+\frac{3}{2}c'^{2}),
 \hspace{2cm}g^{B_{H}l^{+}_il^{-}_i}_{A}=\frac{g'}{2s'c'}(-\frac{1}{5}+\frac{1}{2}c'^{2}),\\
 \nonumber
g^{B_{H}\overline{u_i}u_i}_{V}&=&\frac{g'}{2s'c'}(2y_u+\frac{17}{15}-\frac{5}{6}c'^2),
 \hspace{1.5cm}g^{B_{H}\overline{u_i}u_i}_{A}=\frac{g'}{2s'c'}(\frac{1}{5}-\frac{1}{2}c'^2),\\ \nonumber
g^{B_{H}\overline{d_i}d_i}_{V}&=&\frac{g'}{2s'c'}(2y_u+\frac{11}{15}+\frac{1}{6}c'^2),
 \hspace{2cm}g^{B_{H}\overline{d_i}d_i}_{A}=\frac{g'}{2s'c'}(-\frac{1}{5}+\frac{1}{2}c'^{2}),\\
g^{B_{H}\overline{t}t}_{V}&=&\frac{g'}{2s'c'}(2y_u+\frac{17}{15}-\frac{5}{6}c'^{2}-\frac{1}{5}x_L),
 \hspace{0.6cm}g^{B_{H}\overline{t}t}_{A}=\frac{g'}{2s'c'}(\frac{1}{5}-\frac{1}{2}c'^{2}-\frac{1}{5}x_L).
\end{eqnarray}
$l_i, u_i, d_i$ denote $(e,\mu,\tau),~~(u,c),~~(d,s,b)$,
respectively.  We also define
$x_L=\frac{\lambda_1^2}{\lambda_1^2+\lambda_2^2}$, and $x_{L}$ is
the mixing angle parameter between the SM top quark t and the
vector-like quark T, in which $\lambda_{1}$ and $\lambda_{2}$ are
the Yukawa coupling parameters. As we can see above, the gauge
invariance alone can not unambiguously fix all the $U(1)$
hypercharge values, and the two parameters $y_e$ and $y_u$ are
undetermined. If one requires that the $U(1)$ hypercharge
assignments be anomaly free, they can be fixed as
$y_e=\frac{3}{5}, y_u=-\frac{2}{5}$. In the following calculation,
we take $y_e=\frac{3}{5}, y_u=-\frac{2}{5}$ as an example. In this
case, we can see that an apparently
 special point at $c'=\sqrt{2/5}$ exists where all the
 couplings of the gauge boson $B_{H}$ to light fermion pair vanish. An
 exception is the coupling of $B_{H}$ to the top quark pair, and an additional
 term is attributed to such coupling.

In the LH model, the custodial $SU(2)$ global symmetry is
explicitly broken, which can generate large
 contributions to the electroweak observables. In the early study, global fit
 to the experimental data puts rather severe constraints on the $f>4$ TeV
 at $95\%$ C.L\cite{earlylimit}. However, their analyses are based on a
 simple assumption that the SM fermions are charged only under $U(1)_1$. If the SM fermions are
charged under $U(1)_{1} \times U(1)_{2}$, the bounds become
relaxed. The substantial parameter space allows $f=1\sim2$ TeV,
with $c=0\sim0.5$, $c'=0.62\sim0.73$\cite{newlimit}.

Due to the existence of $B_Hf\bar{f}$ couplings, $B_H$ can be
produced associated with fermion pair via $\gamma \gamma$
collision. The Feynman diagrams for the processes are shown in
Fig.1, in which the cross diagrams with the interchange of the two
incoming photons are not shown.

\begin{figure}[t]
\begin{center}
\epsfig{file=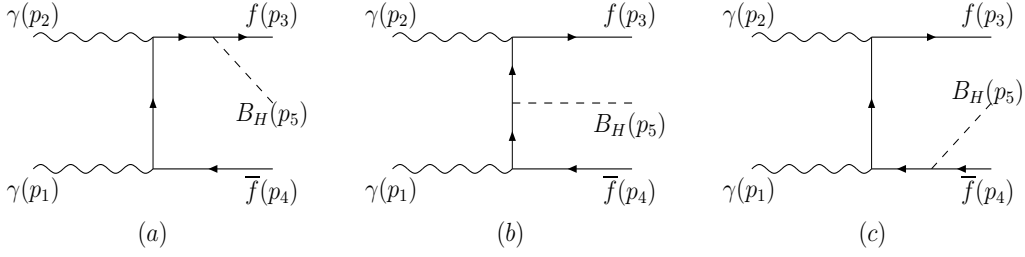,width=450pt,height=700pt} \vspace{-20.5cm}
\caption{\small The Feynman diagrams of the processes
$\gamma\gamma\rightarrow f\bar{f}B_{H}$ in the LH model.}
\label{fig1}
\end{center}
\end{figure}
The amplitudes for the processes are given by
\begin{eqnarray}
M^{a}_{f}&=&iG(p_{3}+p_{5},m_{f})G(p_{1}-p_{4},m_{f})\cdot\\
\nonumber
&&\overline{u}_{f}(p_{3})\eslash(p_{5})g^{B_{H}f\overline{f}}(\pslash_{3}+\pslash_{5}+m_{f})\eslash(p_{2})
g^{\gamma
f\overline{f}}(\pslash_{1}-\pslash_{4}+m_{f})\eslash(p_{1})g^{\gamma
f\overline{f}}v_{f}(p_{4}),\\ \nonumber
M^{b}_{f}&=&iG(p_{3}-p_{2},m_{f})G(p_{1}-p_{4},m_{f})\cdot\\
\nonumber &&\overline{u}_{f}(p_{3})\eslash(p_{2})g^{\gamma
f\overline{f}}(\pslash_{3}-\pslash_{2}+m_{f})\eslash(p_{5})
g^{B_{H}
f\overline{f}}(\pslash_{1}-\pslash_{4}+m_{f})\eslash(p_{1})g^{\gamma
f\overline{f}}v_{f}(p_{4}),\\ \nonumber
M^{c}_{f}&=&iG(p_{3}-p_{2},m_{f})G(p_{4}+p_{5},m_{f})\cdot\\
\nonumber &&\overline{u}_{f}(p_{3})\eslash(p_{2})g^{\gamma
f\overline{f}}(\pslash_{3}-\pslash_{2}+m_{f})\eslash(p_{1})g^{\gamma
f\overline{f}}(-\pslash_{4}-\pslash_{5}+m_{f})\eslash(p_{5})g^{B_{H}
f\overline{f}}v_{f}(p_{4}).
\end{eqnarray}
The amplitudes of the diagrams with the interchange of two
incoming photons can be directly obtained by interchanging
$p_{1},p_{2}$ in above amplitudes. Where
$G(p,m)=\frac{1}{p^{2}-m^{2}}$ is the propagator of the particle.

 \indent With the above amplitudes, we can directly obtain the production cross
sections $\hat{\sigma}(\hat{s})$ for the subprocesses
$\gamma\gamma\rightarrow f\bar{f}B_{H}$ and the total cross
sections at the $e^+e^-$ linear collider can be obtained by
folding $\hat{\sigma}(\hat{s})$ with the photon distribution
function $F(x)$ which is given in Ref.\cite{distribution},

\begin{eqnarray}
\sigma_{tot}(s)=\int^{x_{max}}_{x_{min}}dx_{1}\int^{x_{max}}_{x_{min}
x_{max}/x_1}dx_{2} F(x_{1})F(x_{2})\hat{\sigma}(\hat{s}),
\end{eqnarray}
where $s$ is the c.m. energy squared for $e^+e^-$. The
subprocesses occur effectively at $\hat{s}=x_1x_2s$, and $x_i$ are
the fractions of the electron energies carried by the photons. The
explicit form of the photon distribution function $F(x)$ is
\begin{eqnarray}
\displaystyle F(x)=\frac{1}{D(\xi)}\left[1-x+\frac{1}{1-x}
-\frac{4x}{\xi(1-x)}+\frac{4x^2}{\xi^2(1-x)^2}\right],
\end{eqnarray}
with
\begin{eqnarray}
\displaystyle D(\xi)=\left(1-\frac{4}{\xi}-\frac{8}{\xi^2}\right)
\ln(1+\xi)+\frac{1}{2}+\frac{8}{\xi}-\frac{1}{2(1+\xi)^2},
\end{eqnarray}
and
\begin{eqnarray}
\xi=\frac{4E_0\omega_{0}}{m^{2}_{e}}.
\end{eqnarray}
$E_0$ and $\omega_0$ are the incident electron and laser light
energies, and $x=\omega/E_0$. The energy $\omega$ of the scattered
photon depends on its angle $ \theta $ with respect to the
incident electron beam and is given by
\begin{eqnarray}
\omega=\frac{E_{0}(\frac{\xi}{1+\xi})}{1+(\frac{\theta}{\theta_{0}})^{2}}.
\end{eqnarray}
Therefore, at $\theta =0,~\omega=E_{0}\xi/(1+\xi)=\omega_{max}$
is the maximum energy of the backscattered photon, and
 $x_{max}=\frac{\omega_{max}}{E_{0}}=\frac{\xi}{1+\xi}$.

 To avoid unwanted $e^+e^-$ pair production from the collision between
the incident and back-scattered photons, we should not choose too
large $\omega_0$. The threshold for $e^+e^-$ pair creation is
$\omega_{max}\omega_{0} > m^{2}_{e}$, so we require
$\omega_{max}\omega_{0} \leq m^{2}_{e}$. Solving
$\omega_{max}\omega_{0} = m^{2}_{e}$, we find
\begin{eqnarray}
\xi=2(1+\sqrt{2})=4.8.
\end{eqnarray}
For the choice $\xi=4.8,$ we obtain $x_{max}=0.83$ and
$D(\xi_{max})=1.8.$ The minimum value for $x$ is determined by the
production threshold
\begin{eqnarray}
x_{min}=\frac{\hat{s}_{min}}{x_{max}s},~~~\hat{s}_{min}=(2M_{f}+M_{B_{H}})^{2}.
\end{eqnarray}

Here we have assumed that both photon beams and electron beams are
unpolarized. We also assume that, the number of the backscattered
photons produced per electron is one.

\section{ Numerical results and conclusions}

In our numerical calculations, we take the SM input parameters as:
$m_{e}=0$,
 $m_{\mu}=0.106$ GeV, $m_{\tau}=1.77$ GeV, $m_{u}=0.002$ GeV,
 $m_{d}=0.005$ GeV, $m_{c}=1.25$ GeV, $m_{s}=0.1$ GeV, $m_{t}=174.2$ GeV,
 $m_{b}=4.7$ GeV, $M_Z=91.187$ GeV, $v=246$ GeV, $s^{2}_{W}=0.23$\cite{data}. Another SM parameter,
 the electromagnetic fine structure constant $\alpha_e$, can be fixed at a certain energy scale by calculating from
the simple QED one-loop evolution formula with the boundary value
$\alpha=1/137.04$\cite{Donoghue}. There are still four free
parameters ($f,c,c',\sqrt{s}$) involved in the cross sections,
except for the production mode $t\bar{t}B_H$ with an extra
parameter $x_{L}$. Because the mixing paramter $c$ only has a
little effect on the $B_H$ mass, the cross sections are
insensitive to $c$ and we fix $c=0.3$ as an example. The
influences of parameters $c',f,\sqrt{s},x_L$ on the cross sections
are shown in Figs.2-4, and there we also take into account the
constraints of electroweak precision data on the parameters, i.e.,
$f=1\sim2$ TeV is allowed for the mixing parameters $c,$ $~ c^{'}$
in the ranges of $0\sim0.5$, $0.62\sim0.73$, respectively.

\begin{figure}[ht]
\vspace*{-1.2cm}
\begin{tabular}{cc}
\scalebox{1.68}{\epsfig{file=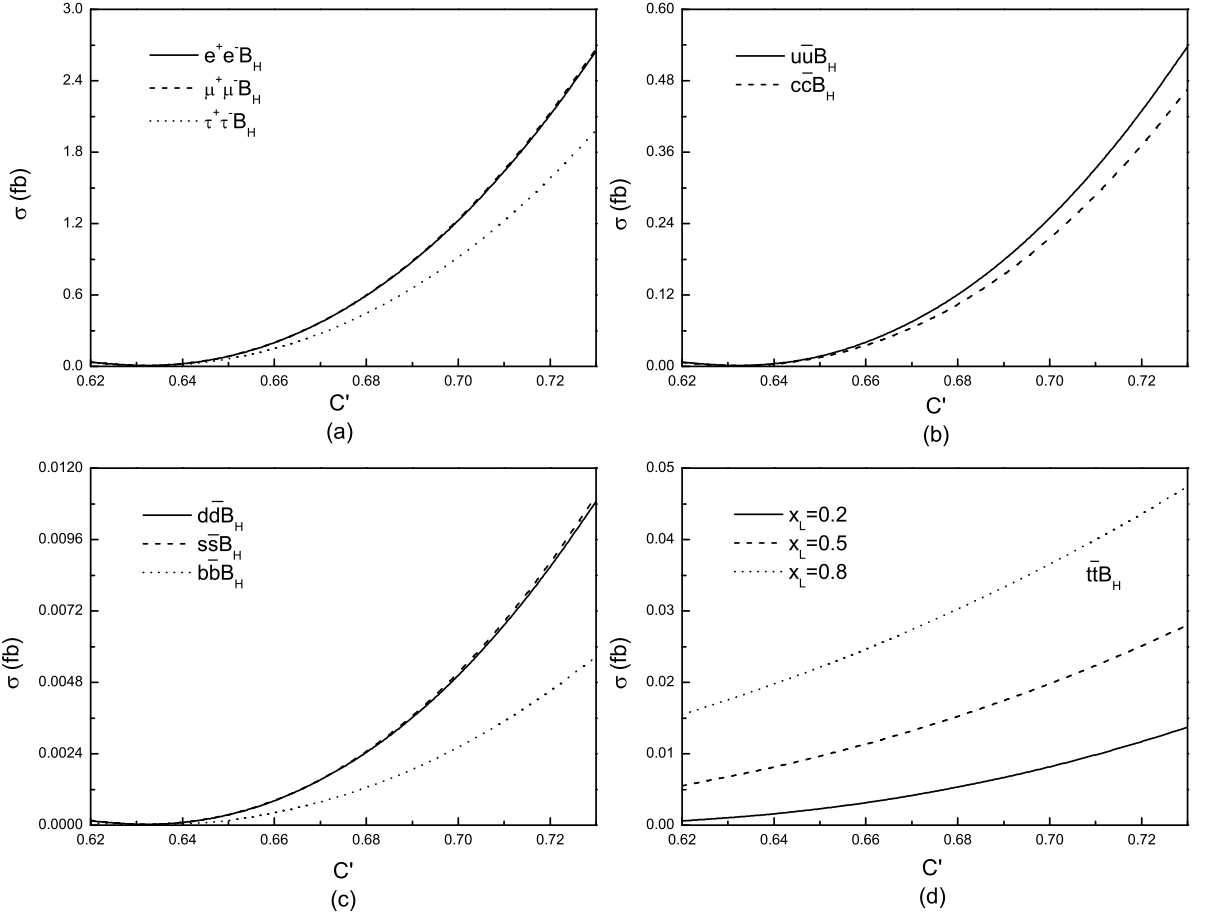}}\\
\end{tabular}
\vspace*{-1.5cm} \caption{\small The production cross sections of
the processes $\gamma\gamma\rightarrow f\bar{f}B_{H}$ as a
function of the mixing parameter $c'$, with $f=2$ TeV, $c = 0.3$,
and $\sqrt{s}$ = 1.5 TeV. Fig.2(a) for the associated lepton pair
productions, Fig.2(b) for the associated light up-type quark pair
productions, Fig.2(c) for the associated down-type quark pair
productions, Fig.2(d) for the associated heavy top pair production
with $x_L=0.2,~0.5,~0.8$, respectively.}
\end{figure}

\indent As we know, $c'$ has strong influence on the couplings
$B_Hf\bar{f}$, and also effects the $B_H$ mass. So the cross
sections should be very sensitive to $c'$. The curves shown in
Fig.2 are the cross sections as a function of $c'$. We can see
that the cross sections sharply drop to zero when $c'$ approaches
$\sqrt{\frac{2}{5}}$ except for the $t\bar{t}B_H$ production.
 This is because the couplings
$B_{H}f\bar{f}$, except for $B_Ht\bar{t}$, are proportional to
$c'^{2}-\frac{2}{5}$(with $y_e=\frac{3}{5}, y_u=-\frac{2}{5}$ for
the anomaly free case) and these couplings become decoupled when
$c'=\sqrt{\frac{2}{5}}$. While, for $c'>\sqrt{\frac{2}{5}}$, the
cross sections increase sharply with $c'$ increasing. The typical
orders of magnitude of the cross sections are: $O(1)$ fb for
$l^+l^-B_H$ productions, $O(10^{-1})$ fb for
$u\bar{u}(c\bar{c})B_H$ productions, $O(10^{-3})$ fb for
$d\bar{d}(s\bar{s},b\bar{b})B_H$ productions, $O(10^{-2})$ fb for
$t\bar{t}B_H$ production. The significant differences of the cross
sections among the associated lepton pair productions, associated
up-type quark pair productions, and associated down-type quark
pair productions are mainly caused by the different coupling
strengths of $f\bar{f}B_H$. The couplings of $B_H$ to lepton pair
are the strongest among $f\bar{f}B_H$, and the couplings of $B_H$
to up-type quark pair are much stronger than those of $B_H$ to
down-type quark pair. The large mass of the top quark sharply
depresses the phase space of the $t\bar{t}B_H$ production which
makes the cross section of the $t\bar{t}B_H$ production is much
smaller than those of $u\bar{u}(c\bar{c})B_H$ productions.

\begin{figure}[ht]
\vspace*{-0.8cm}
\begin{tabular}{cc}
\scalebox{1.65}{\epsfig{file=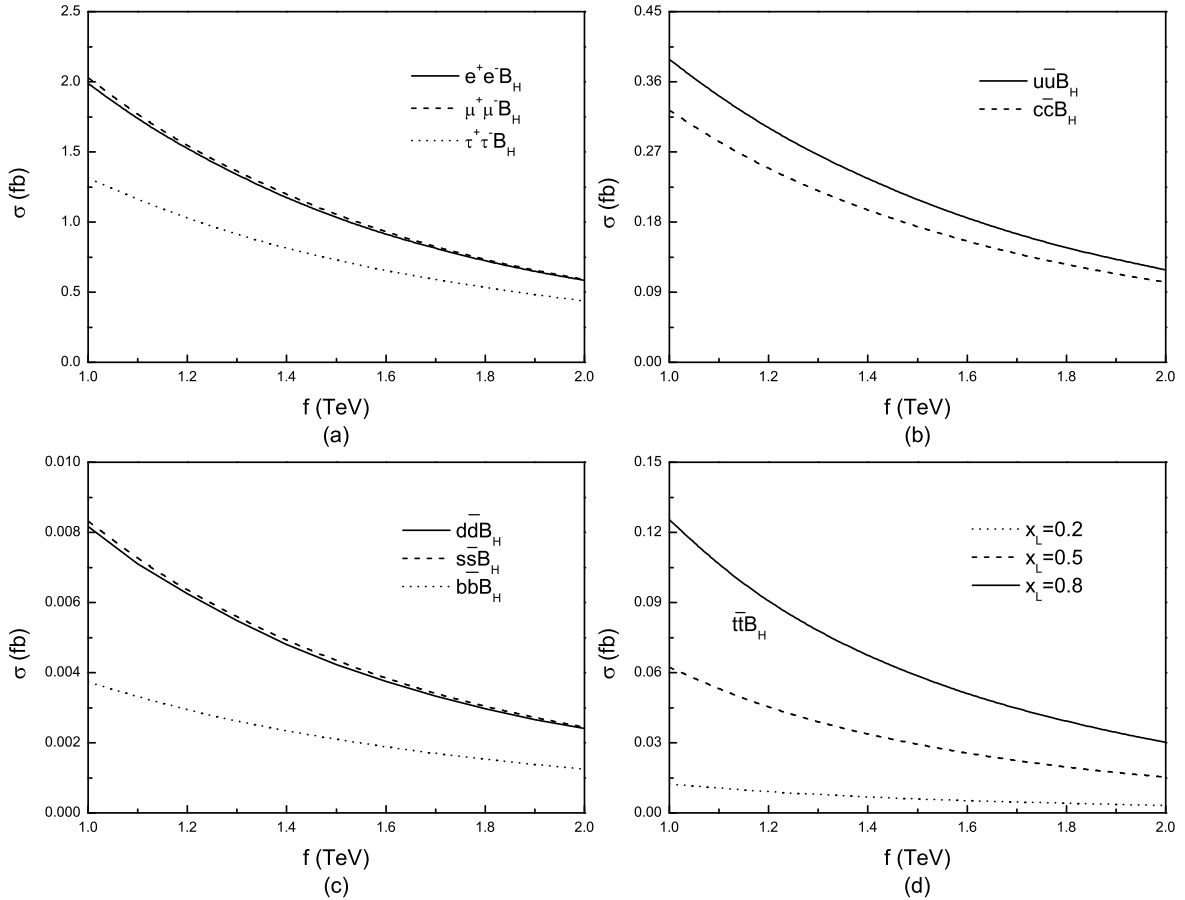}}\\
\end{tabular}
\vspace*{-1.5cm} \caption{\small The production cross sections of
the processes $\gamma\gamma\rightarrow f\bar{f}B_{H}$ as a
function of scale parameter $f$, with $c'=0.68$, $c = 0.3$, and
$\sqrt{s}$ = 1.5 TeV. Fig.3(a) for the associated lepton pair
productions, Fig.3(b) for the associated light up-type quark pair
productions, Fig.3(c) for the associated down-type quark pair
productions, Fig.3(d) for the associated heavy top pair production
with $x_L=0.2,~0.5,~0.8$, respectively.}
\end{figure}

 To see the effect of the scale parameter $f$ on
the cross sections, we plot the cross sections as a function of
$f$ in Fig.3. From the expression of the $B_H$ mass, we can see
that the large value of $f$ can make the $B_H$ mass increase
sharply, meanwhile depresses the phase space significantly. So the
cross sections decrease with the $f$ increasing for the given
$c'$. We also show the plots of the cross sections as a function
of c.m. energy $\sqrt{s}$ in Fig.4. The cross sections
significantly increase with $\sqrt{s}$ increasing at first and
then are insensitive to $\sqrt{s}$. Therefore, the ILC, with c.m.
energy $\sqrt{s}=300-1500$ GeV, can provide an ideal collision
energy to probe $B_H$ via $\gamma\gamma\rightarrow f\bar{f}B_{H}$.

\begin{figure}[h]
\vspace*{-1.2cm}
\begin{tabular}{cc}
\scalebox{1.65}{\epsfig{file=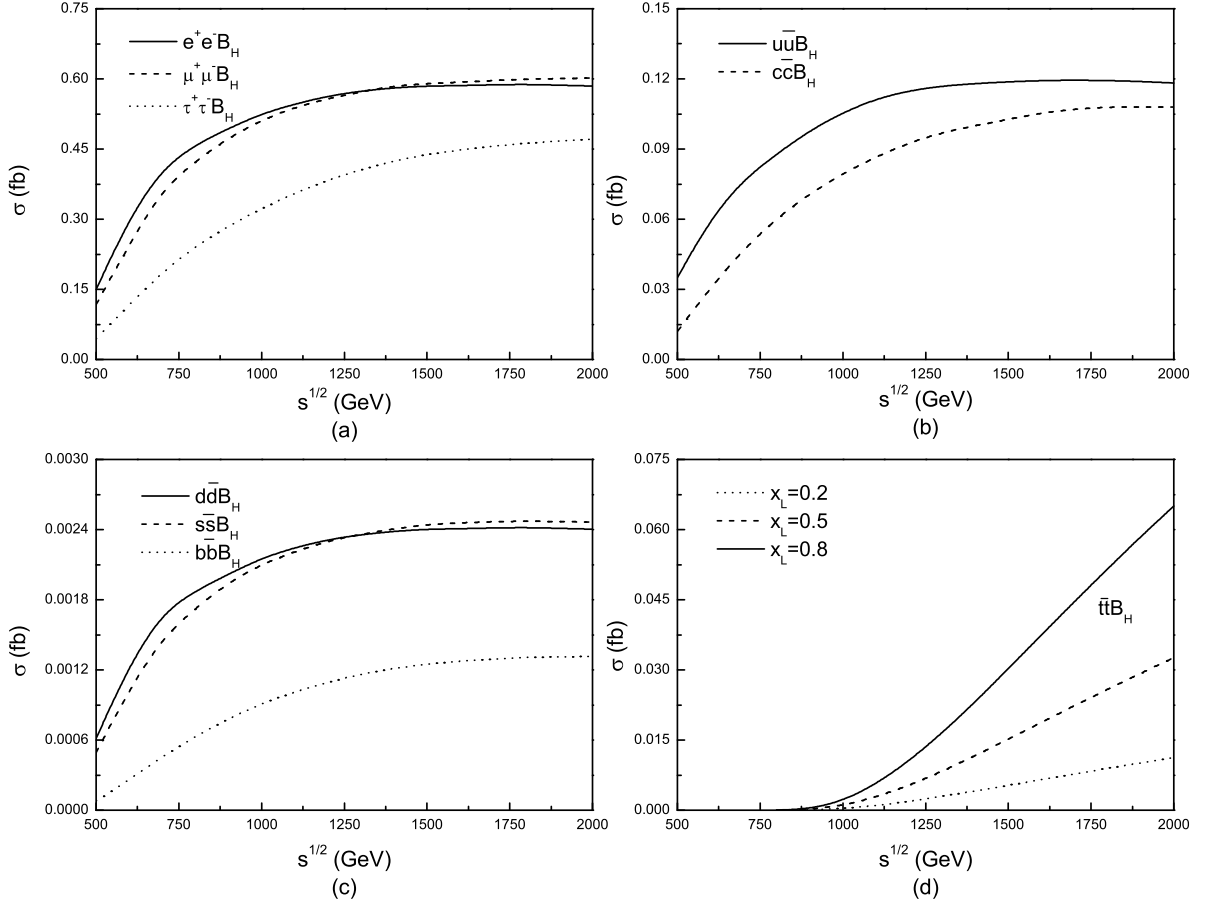}}\\
\end{tabular}
\vspace*{-1.5cm} \caption{\small The production cross sections of
the processes $\gamma\gamma\rightarrow f\bar{f}B_{H}$ as a
function of c.m. energy $\sqrt{s}$, with $c'=0.68$, $c = 0.3$, and
$f= 2$ TeV. Fig.4(a) for the associated lepton pair productions,
Fig.4(b) for the associated light up-type quark pair productions,
Fig.4(c) for the associated down-type quark pair productions,
Fig.4(d) for the associated heavy top pair production with
$x_L=0.2,~0.5,~0.8$, respectively.}
\end{figure}

\indent The integrated luminosity of the ILC can reach $500
fb^{-1}$ within the first four year running. With the cross
sections at the order of $10^{-1}-10^{0}$ fb in most case, there
are enough $B_H$ events can be produced via the production modes
$l^+l^-B_H $. The cross sections of other processes are too small
to probe $B_H$. So we only focus on discussing the production
modes $l^+l^-B_H$. The potential to detect a particle is not only
depended on the event number of the signal produced, but also
depended on the branching ratio of certain decay mode to search
the particle and the efficiency to reconstruct the signal. On the
other hand, the possibility of detection of a particle also
depends critically on the width of the associated resonance, and
wide resonance can be difficult to detect. The decay modes of
$B_H$ have been studied in reference\cite{Han}. For $B_H$, the
parameter spaces where the large decay width would occur are
beyond current search limits in any case. So if $B_H$ would be
produced it can be detected via the measurement of the peak in the
invariant mass distribution of its decaying particles.¡¡The main
decay modes of $B_H$ are $e^+e^-+\mu^+\mu^-+\tau^+\tau^-,
u\bar{u}+c\bar{c}, d\bar{d}+s\bar{s}, W^+W^-,ZH$. The most
interesting decay modes of $B_H$ should be $l'^+l'^-(l'=e,\mu)$.
This is because the leptons $l'$ can be identified easily and the
number of $l'^+l'^-$ background events with such a high invariant
mass is very small. So, a search for a peak in the invariant mass
distribution of $l'^+l'^-$ is sensitive to the presence of $B_H$.
Based on the discussion above, we know that the most interesting
final signals would be $l'^+l'^-l'^+l'^-$ with one $l'^+l'^-$
being reconstructed to $B_H$. These signals that we are interested
in, are not free from the SM backgrounds. In the SM, with
$Z\rightarrow l'^{+}l'^{-}$, the same sufficient final states can
also be produced via the processes $\gamma\gamma\rightarrow
l'^+l'^-Z$($\sim$pb at TeV scale\cite{zll}),
$\gamma\gamma\rightarrow ZZ$($\sim10^2$ fb at TeV scale\cite{zz}).
However, it should be very easy to distinguish $B_H$ from $Z$ when
we look at the $l'^+l'^-$ invariant mass distributions, because
there might exist significantly different $l'^+l'^-$ invariant
mass distributions between $B_H$ and $Z$. Therefore, the
measurement of the $l'^+l'^-$ invariant mass distributions can
greatly depress the background, and the production modes
$\gamma\gamma\rightarrow l'^+l'^-B_H$ with $B_H$ decaying to
$l'^+l'^-$ would open an ideal window to detect $B_H$ with clean
background. As we know, the decay branching ratios of $l'^+l'^-$
approach zero when $c'$ is near $\sqrt{\frac{2}{5}}$. In this
case, one could not search $B_H$ via its leptonic decay modes, and
the bosonic decay modes $W^+W^-, ZH$ would play an important role
in searching $B_H$. The decay branching ratios of these bosonic
decay modes are significant with $c'$ near $\sqrt{\frac{2}{5}}$,
and one can assume enough $W^+W^-$ and $ZH$ signals to be produced
with high luminosity. For the decay mode $B_H\rightarrow W^+W^-$,
the main SM background should arises from $\gamma\gamma\rightarrow
W^+W^-Z$($\sim$pb at TeV scale\cite{zz}) with Z decaying to
$l'^+l'^-$. But the existence of a narrow peak in the $W^+W^-$
invariant mass distribution for the signal can help us to
distinguish $B_H$ from the huge background, so the typical signal
can also be obtained via the decay mode $W^+W^-$. Another
interesting decay mode of $B_H$ is $ZH$ which involves the
off-diagonal coupling $HZB_H$ and the experimental precision
measurement of such off-diagonal coupling is more easier than that
of diagonal coupling. So, the decay mode $ZH$ would provide a
better way to verify the crucial feature of quadratic divergence
cancellation in Higgs mass. Furthermore, such signal would provide
crucial evidence that an observed new gauge boson is of the type
predicted in the little Higgs models. For $B_H\rightarrow ZH$, the
main interesting final states for the production mode
$l'^+l'^-B_H$ should be $l'^+l'^-l^+l^-b\bar{b}$. Two $b$ jets
reconstruct to the Higgs mass and $l^+l^-$ pair reconstructs to
the Z mass. The main SM production processes via the
$\gamma\gamma$ collision can not produce the same final
states\cite{zz},
 so the SM backgrounds for such signal are also very clean.

 \indent In summary, the new gauge bosons in the
 LH model are crucial ingredients for the model. Among these gauge bosons, the
$U(1)$ gauge boson $B_H$ with the mass in the range of hundreds
GeV is the lightest one and  might provide an early signal of the
LH model at the ILC. With the realization of photon-photon
collision at the ILC, the new gauge boson $B_H$ can be produced
via the processes $\gamma\gamma\rightarrow f\bar{f}B_{H}$ which
are studied in this paper. We find that there are the following
features for these processes: (i)The cross sections of these
processes are sensitive to the parameters $f, c'$. (ii)Due to the
large couplings, the cross sections of the processes
$\gamma\gamma\rightarrow l^+l^-B_{H}(l=e,~\mu,~\tau)$ are within
$10^{-1}-10^0$ fb in most parameter spaces allowed by the
electroweak precision data which are the largest among those of
$\gamma\gamma\rightarrow f\bar{f}B_{H}$. With the high luminosity
at the ILC, the sufficient events can be produced to detect $B_H$
via $\gamma\gamma\rightarrow l^+l^-B_{H}$. Specially, the
processes $\gamma\gamma\rightarrow l'^+l'^-B_{H}(l'=e,~\mu)$ could
provide a good chance to detect $B_H$ because the leptons $l'$ can
be easily identified. (iii)For the other processes except
$\gamma\gamma\rightarrow l^+l^-B_{H}$, the cross sections are too
small to detect gauge boson $B_H$. (iv)In most case, the most
interesting decay modes of $B_H$ should be $l'^+l'^-$, and a
search for a peak in the $l'^+l'^-$ invariant mass distributions
are sensitive to the presence of $B_H$ with clean background. When
$c'$ is near $\sqrt{\frac{2}{5}}$, the decay modes $W^+W^-, ZH$
would complement the search for $B_H$. Therefore, the photon
collider realized at the ILC can provide more opportunities to
probe $B_H$ and test LH model.

\end{document}